# Medium-range structural order in amorphous arsenic

Yuanbin Liu[1], Yuxing Zhou[1], Richard Ademuwagun[1], Luc Walterbos[2], Janine George[2,3], Stephen R. Elliott[4], Volker L. Deringer[1]*

[1]*Inorganic Chemistry Laboratory, Department of Chemistry, University of Oxford, Oxford OX1 3QR, UK*

[2]*Materials Chemistry Department, Federal Institute for Materials Research and Testing (BAM), Berlin, Germany*

[3]*Institute of Condensed Matter Theory and Solid-State Optics, Friedrich Schiller University Jena, Jena, Germany*

[4]*Physical and Theoretical Chemistry Laboratory, Department of Chemistry, University of Oxford, Oxford OX1 3QZ, UK*

* volker.deringer@chem.ox.ac.uk

**Medium-range order (MRO) is a key structural feature of amorphous materials, but its origin and nature remain elusive. Here, we reveal the MRO in amorphous arsenic ($a$-As) using advanced atomistic simulations, based on machine-learned potentials derived using automated workflows. Our simulations accurately reproduce the experimental structure factor of $a$-As, especially the first sharp diffraction peak (FSDP), which is a signature of MRO. We compare and contrast the structure of $a$-As with that of its lighter homologue, red amorphous phosphorus ($a$-P), identifying the dihedral-angle distribution as a key factor differentiating the MRO in both. The pressure-dependent structural behaviors of $a$-As and $a$-P differ as well, which we link to the interplay of ring topology and structural entropy. We finally show that the origin of the FSDP is closely correlated with the size and spatial distribution of voids in the amorphous networks. Our work provides fundamental insights into MRO in an amorphous elemental system, and more widely it illustrates the usefulness of automation for machine-learning-driven atomistic simulations.**

## INTRODUCTION

The central structural feature of amorphous materials, beyond the short-range order of nearest-neighbor coordination shells, is the medium-range or intermediate-range order (MRO / IRO) at distances of 5–20 Å (*1*). For many years, studies of MRO have advanced our fundamental understanding of amorphous solids, and now they could help to "design" the latter (*2*) based on the correlation of MRO with macroscopic properties (*3*, *4*). For example, enhanced MRO in vapor-deposited $GeO_2$ glass, identified through Raman spectra, was shown to reduce room-temperature internal friction (*5*). In addition, the strengthening of MRO has been linked to increased thermal conductivity in *a*-Si (*6*), *a*-$Ga_2O_3$ (*7*), and *a*-C (*8*). Experimental techniques, such as X-ray or neutron scattering, have been widely used to probe MRO for different systems. In particular, the first sharp diffraction peak (FSDP) in the structure factor, $S(Q)$, has long been regarded as a signature of MRO (*9–11*). However, extracting and interpreting structural information about MRO from experimental scattering data is non-trivial, which is further complicated by its sensitivity to pressure (*12*, *13*) and compositional variations (*14*).

An interesting fundamental question is what similarities or differences exist in the MRO of elemental glasses within the same group of the Periodic Table. In the present work, we will address this question for two group-15 elements, phosphorus (P) and arsenic (As). The crystalline allotropes are now well understood, and P is particularly rich in structures: exhibiting puckered layers in black P, complex tubular structures in violet P, and tetrahedral $P_4$ molecules in white P. The heavier homologue, As, also adopts diverse structures, from gray As, which is isostructural with high-pressure rhombohedral P, to yellow As comprising $As_4$ molecules. In terms of the disordered state, amorphous phosphorus (*a*-P) displays pronounced MRO, characterized by clusters formed primarily of five-membered rings (*13*, *15*, *16*). In contrast, the MRO of amorphous arsenic (*a*-As) and its structural relationships with the crystalline allotropes remain to be explored.

Insights into the structure of amorphous materials are increasingly obtained from molecular-dynamics (MD) simulations which can directly probe MRO at the atomic scale, complementing experiments. By topological and geometrical analyses of MD trajectories, such as primitive ring statistics (*17*), Voronoi tessellation (*18*), or persistent homology (*19*), one can identify structural motifs contributing to MRO. However, the reliability of this approach strongly depends on the accuracy of the force predictions used to drive the simulations, and modelling at the level of density-functional theory (DFT) has been limited to rather small simulation-cell sizes (*20*). Recent, rapid progress in machine-learned interatomic potentials (MLIPs) (*21–23*) has unlocked simulations of amorphous materials at large length-scales and for long simulation times. While developing MLIPs for amorphous materials has traditionally required extensive domain expertise and manual data curation (*24*, *25*), the emergence of automated workflows is now poised to substantially accelerate the construction of MLIPs (*26–29*). This means that simulations that would previously have required the careful construction of a hand-crafted MLIP model can now be carried out much more quickly than before.

Here, we show how random structure searching (RSS), implemented within automated workflows (*29–31*) and refined by a few iterations of MD, can generate a high-quality training dataset and MLIP model for *a*-As with very moderate computational effort. We first validate our MLIP by quantitatively comparing the computed structure factor of *a*-As—and especially the FSDP—with experimental data. We then use ML-driven MD simulations to uncover the MRO in *a*-As and to draw a comparison to recent studies of *a*-P (*15*, *16*). Specifically, by examining the dihedral-angle distributions in *a*-As and *a*-P, we identify geometric motifs responsible for differences in the MRO of both elements. We also reveal unexpected differences in the high-pressure structural evolution of *a*-As and *a*-P. Finally, we show that voids in *a*-As can explain the origin of the FSDP. Our work contributes to a deeper understanding of MRO in amorphous materials and shows how such an understanding can be obtained with the help of automated atomistic machine learning.

## RESULTS

**Machine-learned potentials for arsenic**

The starting point for our studies was to use the Gaussian Approximation Potential (GAP) framework (*22*, *32*, *33*) to iteratively explore and sample the potential-energy surface of elemental As, using small cells of up to 24 atoms (Fig. 1A). Previous studies have shown that such GAP-driven random structure searching (GAP-RSS) can efficiently capture diverse atomic environments and yield robust potentials at low computational cost (*29*, *30*, *34*). Here, the RSS processes were carried out automatically using the `autoplex` framework we have developed recently (*29*). After accumulating a dataset of 1,500 RSS-generated configurations, we refitted the potential-energy surface using the MACE architecture (*35*). The rationale for doing so is that GAP is data-efficient, and thus particularly suitable for initial RSS, whereas MACE achieves higher accuracy once sufficient data are available; a similar staged approach was used previously to build MLIPs for graphene oxide (*36*). We evaluated our initial MACE model, trained on the pure RSS dataset, through melt-quench MD simulations yielding structural models of *a*-As at 300 K. The $S(Q)$ data calculated for those models were then compared with two experimental datasets: one reported by Smith et al. (*37*), constructed from combined X-ray (*38*) and neutron data, and one reported by Bellissent and Tourand (*39*), obtained from neutron scattering measurements. Although the comparison shows overall agreement with both sets of experimental data, the simulated FSDP height is underestimated (Fig. 1B).

To enhance the representation of MRO in the training data, we next carried out iterative MD refinements in larger, 216-atom cells, using the RSS-derived MACE model as the starting point. Each iteration involved simulations across 5 distinct densities, from 4.14 to 5.34 g cm$^{-3}$ at intervals of 0.3 g cm$^{-3}$. We selected this range to encompass the experimental density of *a*-

As at ambient pressure (4.74 g cm$^{-3}$) (*37*). After four iterations, the refined potential yielded a structure in excellent agreement with the experimental *S*(*Q*) data, notably reproducing the FSDP height with a relative error below 1% (Fig. 1C). We note that the additional MD-based refinement mainly enhanced the intensity of the predicted FSDP, while the rest of the curve is almost identical to that predicted by the purely RSS-trained model (Fig. S1).

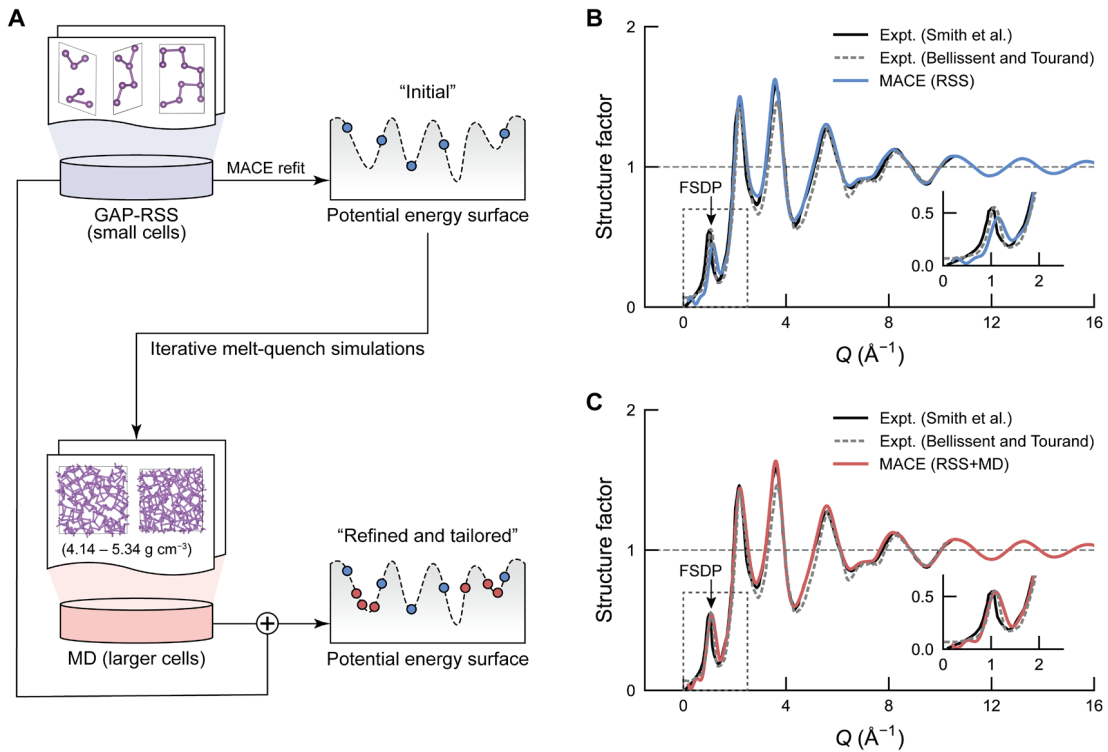

**Fig. 1 Machine-learning-driven simulations of *a*-As.** (**A**) Schematic workflow for developing MLIP models for As. In the first stage, the `autoplex` framework was used to perform GAP-RSS on small cells, generating initial training data, to which we then fitted a graph-based MACE model. In the second stage, the model was iteratively refined by MD melt-quench simulations on larger (216-atom) cells across a density range of 4.14–5.34 g cm$^{-3}$. (**B–C**) Simulated structure factors from the two MLIP variants as compared to experimental data digitized from Smith et al. (*37*) as well as Bellissent and Tourand (*39*). Panel B shows results for the initial MACE potential trained on RSS data only, while panel C shows results from the refined potential that also incorporates MD data. The refined potential improves the agreement with experimental data for the FSDP (arrows and insets in both panels).

To place our results in methodological context, we compared them to those of a state-of-the-art atomistic foundation model, MACE-MPA-0 (*40*), which was designed for broad applicability across diverse chemical systems. While the zero-shot foundation model (i.e., without fine-tuning) qualitatively captures the main features of the experimental structure factor for *a*-As, it overestimates the height of the FSDP while simultaneously underestimating that of the second peak (Fig. S2). This comparison underscores that, while foundation models provide a valuable baseline, achieving high fidelity for amorphous systems still requires a targeted training and refinement strategy. Our present work is concerned with building MLIP models "from scratch", but we mention in passing that the same methodology could provide a pathway for fine-tuning existing foundation models, as relevant training data can be generated with high computational efficiency and minimal human intervention. Specifically, the RSS stage in the present work required < 15,000 CPU core hours, and the total computational cost for both the RSS and MD stages was < 50,000 CPU core hours.

The cost-efficiency of our approach also makes it suited for comparing and benchmarking different exchange–correlation functionals to be used for generating training data. In addition to r$^2$SCAN (*41*), we generated separate MLIP models with TPSS (*42*) and r$^2$SCAN+rVV10 (*43*), respectively, using the same protocol otherwise. The r$^2$SCAN+rVV10 approach augments r$^2$SCAN with a nonlocal van der Waals correlation term, originally designed to improve the treatment of long-range dispersion interactions (*43*). While this extension benefits layered materials (*44*), we find that in the present case of *a*-As it alters medium-range correlations in a way that severely underestimates the intensity of the FSDP (Fig. S3). TPSS is closer to the r$^2$SCAN predictions, but still overestimates the intensity of the second *S*(*Q*) peak (Fig. S3). These results underline how sensitive the simulation of amorphous structures can be to the choice of functional, and our automated framework offers a practical and economic route to address this important challenge. More details on the GAP-RSS procedure, MLIP fitting, and melt–quench protocols are provided in the Methods section.

**Medium-range structural order**

Our main MLIP model was used to generate 2,000-atom models of *a*-As (cell length ≈ 38 Å), and we first probed the short-range structural order, as described by radial distribution functions and bond-angle distributions, at ambient conditions. Figure 2A presents the radial distribution functions, *J*(*r*), for *a*-As. We also include results for an earlier simulation of *a*-P (*15*), in which amorphous structures were obtained using an MLIP (*45*) trained on PBE+MBD data (*46–48*), noting that the latter is a different ground-truth level compared to the present work. To carry out a side-by-side comparison and account for the different bond lengths, the *J*(*r*) data for *a*-P and *a*-As were rescaled to the position of the respective first peak. In this, *a*-P exhibits a slightly narrower first peak and a broader subsequent minimum (approaching zero) than *a*-As, indicating a more uniform bond-length distribution and a more well-defined first coordination shell. The second peak in *a*-P is remarkably sharp and intense, concomitant with a narrow bond-angle distribution (Fig. 2B); in contrast, there is a less strong second-nearest-neighbor peak in *a*-As, and greater angular disorder. A well-defined third main *J*(*r*) peak for *a*-P and *a*-As shows a clear hallmark of MRO. We also provided a direct comparison between the calculated *J*(*r*) and experimental data for *a*-As, further demonstrating the accuracy of our MLIP (Fig. S4).

One important indicator of MRO is the distribution of primitive rings (Fig. 2C). Notably, 5-membered rings dominate the network topology of both *a*-P and *a*-As. This can be attributed to the fact that both materials exhibit bond-angle distributions consistent with 5-membered ring formation (Fig. S5). However, *a*-P contains a significantly higher proportion of 5-membered rings than *a*-As, consistent with the greater probability density for their requisite bond angles (Figs. 2B–C, Fig. S5). Interestingly, 5-membered rings are also the primary structural motif in violet (Hittorf's) (*49*) and fibrous (Ruck's) phosphorus (*50*). Compared to *a*-P, for which the proportion of large rings rapidly diminishes beyond *n* = 6, *a*-As contains a larger proportion of extended-size rings (*n* > 6) (Fig. 2C).

Another signature of MRO can be the formation of fragment clusters. Such clusters composed of 5-membered rings, following Baudler's rules (*51*) were previously reported in *a*-P (*26*) and are also observed here in *a*-As (Fig. 2D). Typical examples of these clusters range from motifs X2]X3[X2 (two fused 5-membered rings sharing three atoms) and X3]X2[X3 (two fused 5-membered rings sharing two atoms) to cage-like X8, X9, and X10 fragments (X = P, As). However, *a*-As exhibits a lower abundance of these clusters (Fig. 2D), which can be partly attributed to the limited number of 5-membered rings in *a*-As, which reduces the available building blocks for cluster formation.

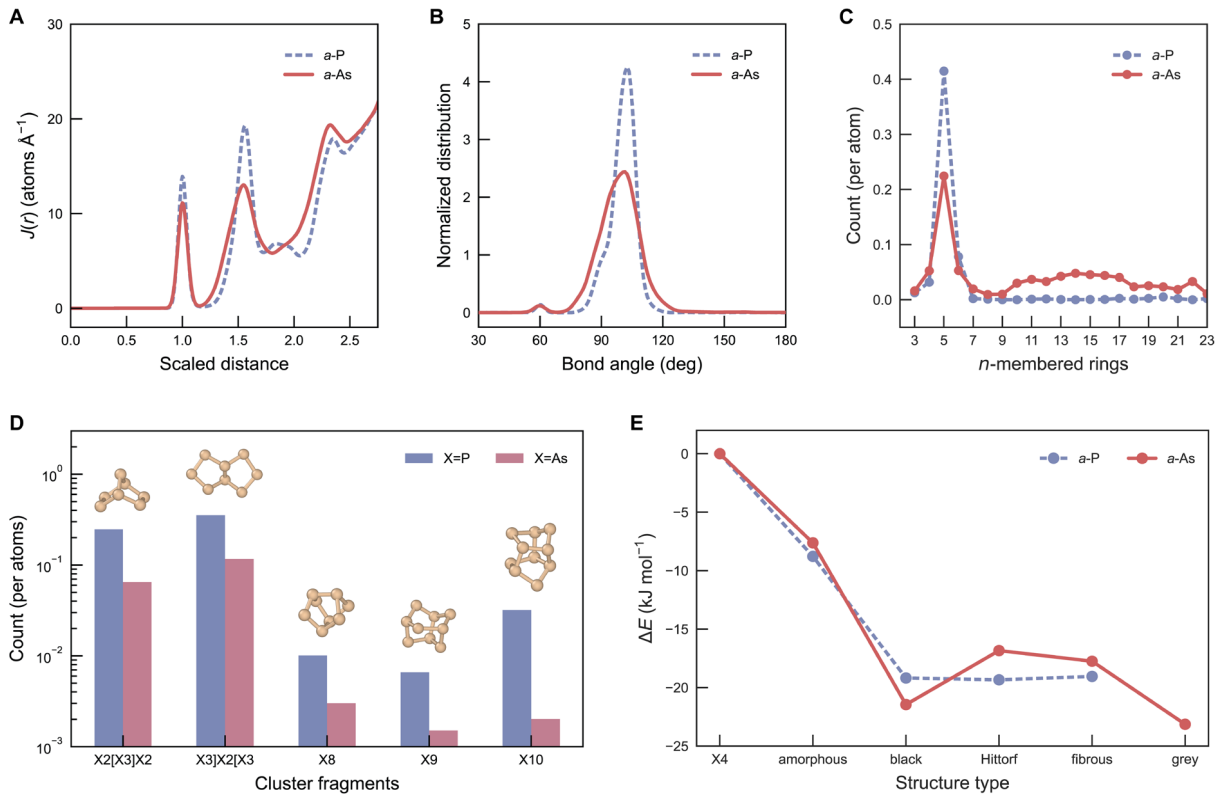

**Fig. 2 Structural and energetic properties of *a*-As at ambient pressure, compared with *a*-P.** (**A**) Radial distribution functions, $J(r)$, of *a*-P [dashed purple line; structure from Ref. (*16*)] and *a*-As (solid red line). The distance, $r$, is scaled by the first peak position in $J(r)$. (**B**) Bond-angle distribution for *a*-P and *a*-As. The plot shows the probability-density functions of the total bond-angle distributions calculated for all atoms in the systems. (**C**) Distribution of primitive rings. (**D**) Counts of cluster fragments. It can be observed that the proportions of different types of clusters are higher in *a*-P than in *a*-As. (**E**) The energetics of arsenic and phosphorus allotropes, including the well-known $X_4$ phases (yellow for X = As, white for X = P, used as the reference state in both cases), as well as the hypothetical violet (Hittorf-type) and fibrous forms.

The hypothetical Hittorf-type and fibrous As structures are derived from the corresponding phosphorus phases (*16*) and subsequently relaxed using DFT. Note that the energetics of all As forms were computed at the r$^2$SCAN+rVV10 level of theory, whereas the P data were taken from Ref. (*16*), which used the HSE06+MBD functional (*47, 48, 52*) for the energy calculations.

Continuing our comparison of both elements, we examined the energetic stability of crystalline and amorphous phases of As. In addition to the gray, black, and yellow allotropes, we also included hypothetical As phases isostructural with violet (Hittorf's) and fibrous P, constructed by elemental substitution and subsequent full structural relaxation. We note that violet P can be synthesized from amorphous red P (*53*), but no similar synthesis pathway starting from *a*-As has been reported to our best knowledge. In a benchmark comparison of DFT functionals for crystalline forms of As (Table S1), we found that r$^2$SCAN, due to its lack of long-range (van der Waals) interaction description, fails to correctly optimize the structure of layered gray arsenic. In contrast, the r$^2$SCAN+rVV10 functional provides higher accuracy for crystalline structures and was therefore employed here to fully relax all crystalline structures. Given the substantial computational cost, we relaxed the 500-atom *a*-As structures using MLIPs, followed by single-point energy evaluations with r$^2$SCAN+rVV10. This approach allows us to directly compare the energies of amorphous and crystalline phases. Our results show that hypothetical, Hittorf-type and fibrous phases of arsenic are energetically less stable than the known gray and black allotropes (albeit they are more favorable than *a*-As). The gray phase exhibits the highest energetic stability, consistent with previous literature (*54*). White P and yellow As are both higher in energy than their respective amorphous counterparts. Projected Crystal Orbital Hamilton Population (COHP) analysis (*55, 56*) using LOBSTER (*57*) and additional automated postprocessing with LobsterPy (*58*) was carried out to gain additional insight into the electronic structure and chemical-bonding properties of the relevant structures, including *a*-As (Supplementary Text and Figs. S6 and S7).

**Role of dihedral angles in MRO**

Group-15 elements typically adopt threefold-coordinated p-bonding environments. In the crystalline phases of As, atoms are all 3-fold coordinated, whereas in their amorphous counterpart, occasional over-coordinated sites (4- and 5-fold) may occur as structural defects, and the population of these tends to increase under pressure (Fig. S8). However, both *a*-P and *a*-As exhibit almost exclusively 3-fold coordination at ambient pressure, with negligible under- (2-fold) and over-coordination (Fig. S8).

The predominant 3-fold coordination in both crystalline and amorphous phases gives rise to numerous "dumbbell-like" building units, each formed by two connected trigonal units (Fig. 3A). We quantified the torsional geometry of dumbbell units using a defined dihedral angle, $\phi$ (Methods). Rhombohedral gray arsenic exhibits a single dihedral angle of 180°, whereas the orthorhombic black modification is characterized by three discrete $\phi$ values of ±78.7° as well as 180°, consistent with the formation of puckered zigzag chains. In contrast, *a*-As displays a higher structural (torsional) flexibility, with dumbbell dihedral angles spanning a broader, more uniform distribution (Fig. 3B). Interestingly, while *a*-P also shows a higher torsional flexibility than its crystalline counterparts, its dumbbells mostly feature dihedral angles of $|\phi| < 90°$. To explain the uniformity of the dihedral-angle distribution in *a*-As, we explored their relationship with average bond energies (ICOHPs) in Fig. S7B. The weak correlation, with a Spearman coefficient (*59*) of 0.0057, indicates that the bond energy is largely independent of the dihedral angle, resulting in a more uniform angular distribution.

The formation of extended-size rings ($n > 6$) in *a*-As, as opposed to *a*-P, can be attributed to its broader distribution of dumbbell dihedral angles, which enables greater structural flexibility. This finding was further confirmed by analyzing the dihedral-angle distribution bias, which we define as the difference between normalized angle distributions for extended ($n > 6$) and small ($n \leq 6$) rings, respectively (Fig. 3C). This analysis shows that small rings are preferentially formed from dumbbells with $|\phi| < 90°$, while larger rings predominantly include dumbbells with $|\phi| > 90°$ to accommodate their extended geometry. Since *a*-As contains a higher proportion of large-$\phi$ dumbbells than *a*-P (Fig. 3B), it has a greater propensity to form larger rings.

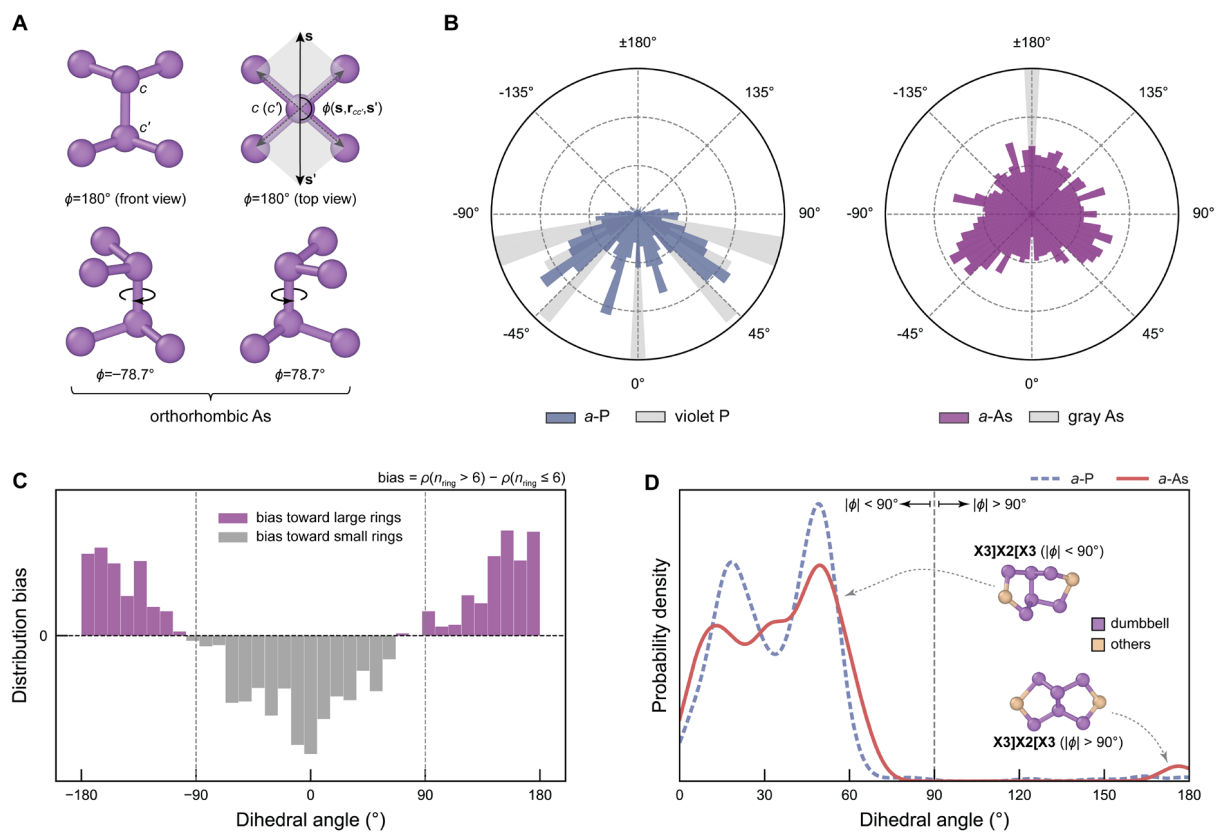

**Fig. 3 Analysis of dumbbell dihedral angles.** (**A**) Illustration of typical "dumbbell" building units in crystalline orthorhombic (black) arsenic. A value of the dihedral angle of $\phi = 180°$ corresponds to an anti-aligned configuration, with the As atoms bonded to each central As positioned opposite to each other along the *c*–*c'* axis. (**B**) Distributions of dihedral angles in *a*-As and *a*-P. (**C**) Distribution bias in dihedral angles between large ($n > 6$) and small ($n \leq 6$) rings in *a*-As. The bias is calculated as the difference between the normalized dihedral-angle distributions of dumbbells in large rings and those in small rings. (**D**) Distributions of dihedral angles in X3]X2[X3 fragment clusters, the most prevalent cluster type, shown as kernel density estimates for *a*-P (dashed purple line) and *a*-As (solid red line).

Regarding the formation of fragment clusters, in addition to the influence of 5-membered rings on the availability of building blocks, the distribution of dumbbell dihedral angles also plays a significant role in determining the difference in cluster populations between *a*-As and *a*-P. Clusters such as X8, X3]X2[X3, X9, and X10 (X = P, As) contain varying numbers of dumbbell units (see insets in Fig. 2D), and these compact clusters largely require dumbbells with $|\phi| < 90°$. The latter is quantitatively illustrated for the dominant cluster type, X3]X2[X3, in Fig. 3D. This figure shows that both materials predominantly utilize dumbbells with $|\phi| < 90°$ to form X3]X2[X3 clusters. However, the structure of *a*-P appears to be more restricted to these small-$\phi$ dumbbells (Fig. 3B), which could explain the higher prevalence of these fragment clusters as compared to *a*-As (cf. Fig. 2D). We emphasize that the above conclusions hold for structures predicted using MLIPs trained with different meta-GGA functionals, including r$^2$SCAN, r$^2$SCAN+rVV10, and TPSS (Fig. S9).

**Evolution of MRO under pressure**

We next explored the evolution of MRO under pressure, which again revealed distinct behaviors in ring and cluster fragment distributions between *a*-P and *a*-As. The evolution of 5-membered rings under pressure reveals a striking contrast between the two materials (Fig. 4A). In *a*-P, their fraction decreases monotonically with compression, consistent with the pronounced weakening of the bond-angle peak characteristic of 5-membered ring geometries (Fig. S5). Conversely, *a*-As exhibits a continuous *increase* in 5-membered rings, suggesting that compression promotes the formation and stabilization of this motif despite increasing angular distortions (Fig. S5).

Figure 4B shows that the population of large rings ($n > 10$) in *a*-P increases with increasing pressure. This is consistent with earlier studies, suggesting that cage-like motifs tend to open under compression, transforming into extended ring structures (*15*). In contrast, the number of large rings in *a*-As remains relatively stable with increasing pressure, exhibiting minimal change, even at elevated pressures (Fig. 4B). To rationalize this difference, we used the information entropy $\mathcal{H}$ (*60*) to quantify the structural diversity encoded in the dihedral-angle distribution, which is given as:

$$\mathcal{H} = \int -\rho(f) \log_2 \rho(f)\, df, \qquad (1)$$

where $\rho(f)$ is the probability density of the dihedral angle, and the logarithm is taken with base 2, suggesting that $\mathcal{H}$ is measured in units of bits. A similar concept was proposed by Schwalbe-Koda et al. who used the information entropy to quantify the completeness of atomistic datasets (*61*). For *a*-P, the increase in extended rings with pressure coincides with an increase in the dihedral-angle entropy (Fig. 4C), indicating an increase in configurational diversity of the dumbbell shapes. This suggests that at higher pressure, the structural flexibility of *a*-P increases, enabling the formation of extended rings. For *a*-As, the dihedral-angle entropy is already high at standard pressure and both the number of extended rings and entropy remain practically unchanged under applied pressure (Fig. 4C). Thus, changes in the ring statistics might be rationalized by the underlying dihedral-angle distributions, but it is clear that the underlying mechanism is different for *a*-As than for *a*-P.

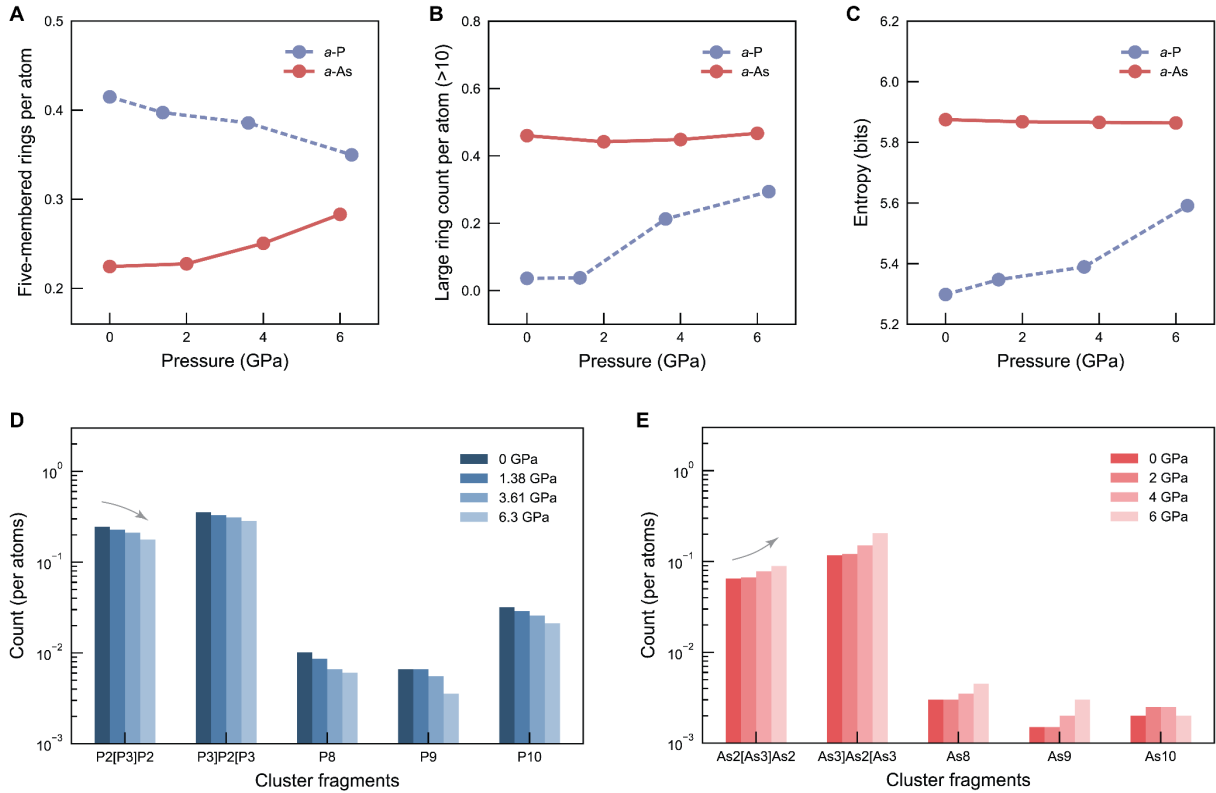

**Fig. 4 Pressure-dependent structural evolution.** (**A**) Count of five-membered rings under pressure for *a*-P and *a*-As. (**B**) Large-ring count per atom ($n > 10$) versus pressure for *a*-P and *a*-As. (**C**) Information (or Shannon) entropy of dihedral-angle distributions (in bits) for *a*-P and *a*-As. Here, the entropy quantifies the breadth and uniformity of the dihedral-angle distribution, reflecting the configurational diversity of local structures. (**D**–**E**) Counts of representative cluster fragments in *a*-P and *a*-As, respectively. The arrows indicate the direction of change with pressure.

Looking at the evolution of cluster fragments under pressure (Figs. 4D–E), the contrasting behavior between *a*-P and *a*-As becomes even more apparent. In *a*-P, increasing pressure results in a clear reduction of all considered fragment clusters, whereas in *a*-As, the abundance of analogous cluster fragments increases under compression. This opposite behavior aligns closely with the changes in 5-membered ring counts in both systems: fewer 5-membered rings with increasing pressure in *a*-P restricts the formation of these compact clusters, while their increase in *a*-As with increasing pressure facilitates cluster assembly (Fig. 4A). Moreover, the enhanced diversity of the dihedral-angle distribution in *a*-P results in a decreased probability of small-angle dihedral angles, which are key to cluster formation, as discussed before (Fig. 3D). As a result, the cluster formation is suppressed in *a*-P as the pressure is increased. The observed structural evolution under pressure emphasizes unexpected contrasting responses of medium-range order between these two chemically similar, yet structurally distinct, amorphous materials.

**Origin of the FSDP**

Like for *a*-P, the structure factor, $S(Q)$, of *a*-As exhibits a pronounced FSDP (Fig. 1C), a well-established signature of MRO in amorphous materials that has previously been linked to the spatial distribution of voids (*9*, *62*). To further investigate its microscopic origin, we examined the pressure dependence of the FSDP, alongside structural features associated with voids (Fig. 5). Figure 5A shows representative atomic structures of *a*-As and *a*-P, with voids visualized as semi-transparent purple and blue regions, respectively. These voids reveal the heterogeneous spatial arrangement of free volume and the distinct structural differences between *a*-As and *a*-P. Specifically, *a*-As shows a local layered arrangement in its void structure, reminiscent of the characteristic layering in crystalline black As, which consists of puckered layers held together by van der Waals interactions (*54*). This structural similarity between amorphous and crystalline As has also been highlighted in previous studies through comparable optical-reflection spectra, which indicate similar coordination environments and bonding arrangements (*63*).

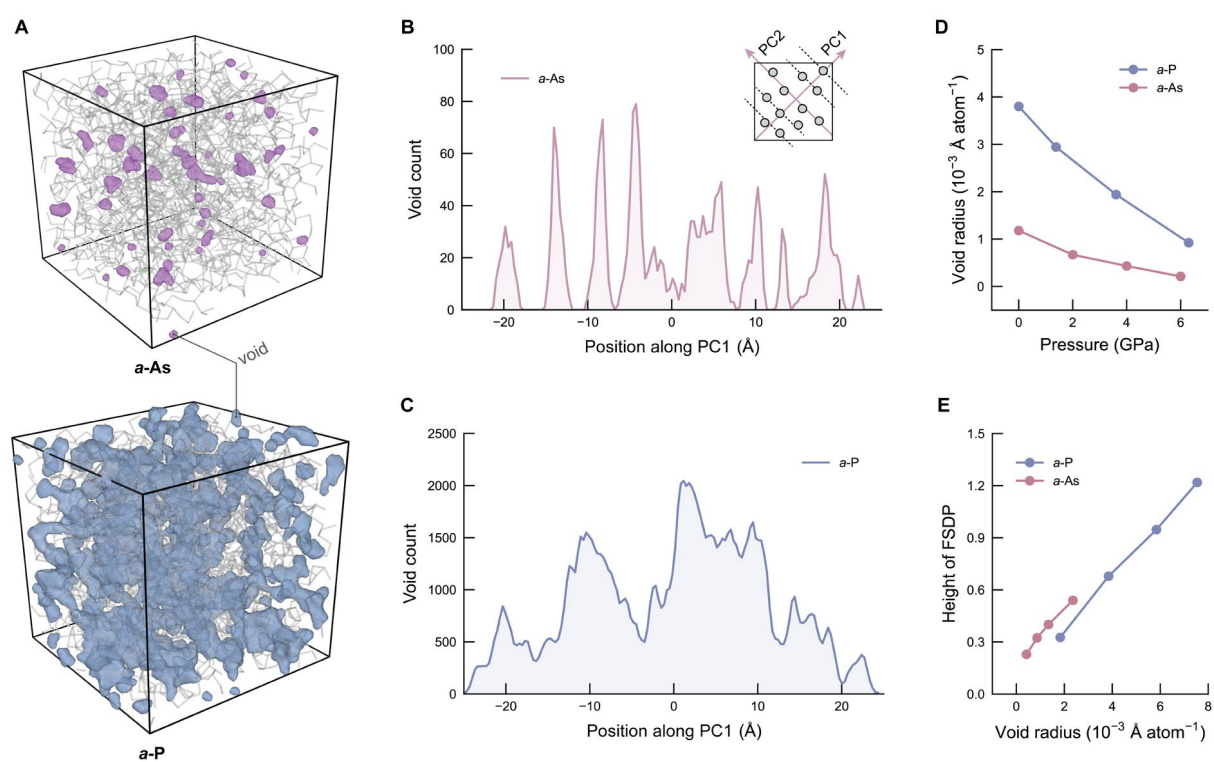

**Fig. 5 Correlation between voids and FSDP intensity.** (**A**) Atomic structures of *a*-As (top) and *a*-P (bottom), with voids visualized as semi-transparent purple and blue regions, respectively. The atomic coordinates of *a*-P were obtained from the literature (*15*). It shows that the void region in *a*-P is much bigger than that in *a*-As. (**B**) Spatial distribution of voids in *a*-As projected along the first principal-component (PC1) axis obtained from principal-component analysis (PCA) on the void coordinates. The inset provides a schematic of PCA. (**C**) Same but for *a*-P. (**D**) Comparison of the average equivalent void radius change under pressure between *a*-As and *a*-P. The equivalent void radius is a normalized metric calculated by determining the radius of a hypothetical sphere whose volume equals the sum of all void clusters, then averaging this value over the total number of atoms in the system. (**E**) Variation of FSDP height with average equivalent void radius, showing a strong linear correlation.

To better understand this behavior, we applied principal-component analysis (PCA) to project the spatial distribution of voids onto a new coordinate system, where each axis is called a principal component, as illustrated in the inset of Fig. 5B. Specifically, the first principal component (PC1) captures the primary spatial direction along which the void density varies most significantly. The distribution of void volumes along PC1 suggests that *a*-As exhibits an anisotropic, layered arrangement of voids, characterized by distinct, directional peaks in the void count (Fig. 5B). We note that the importance of layer-like structural features has been pointed out in an earlier modelling study (*64*). In contrast, the void distribution in *a*-P appears more isotropic, with voids evenly and uniformly dispersed throughout the structure, resulting in a relatively flat profile along PC1 (Fig. 5C). The distributions of voids along the second and third principal components (PC2 and PC3) are shown in Fig. S10.

Moreover, we quantified the pressure dependence of the average equivalent void radius which is defined by:

$$r = (3V/4\pi)^{1/3}/N, \tag{2}$$

where *V* is the sum of all void cluster volumes and *N* denotes the number of atoms in a system. In both materials, the equivalent void radius decreases with increasing pressure (Fig. 5D). The FSDP height exhibits a strong linear correlation with the equivalent void radius, with similar slopes observed in the two systems (Fig. 5E). Together, these results support the interpretation that the FSDP arises from voids in *a*-As, which becomes increasingly disrupted under compression.

**DISCUSSION**

We have shown that, although both P and As prefer similar, 3-fold-bonded local atomic environments, their amorphous modifications exhibit markedly different medium-range structural order. At ambient pressure, *a*-As contains many large rings ($n > 6$), which are practically absent in *a*-P. We attribute this structural divergence primarily to differences in the diversity of dihedral-angle distributions: specifically, the greater torsional flexibility of *a*-As bonds allows more extended ring structures to form. The void structure in *a*-As appears to be layered to some extent, whereas *a*-P exhibits a more isotropic and uniform distribution of voids.

Correspondingly, compact clusters—predominantly built from five-membered rings and dumbbells with dihedral angles less than 90°—are more prevalent in *a*-P. Fewer five-membered rings and a broader dihedral-angle distribution in *a*-As result in fewer compact clusters. Under compression, the covalent network of *a*-P evolves toward increasingly large ring structures with fewer compact clusters (*15*), whereas *a*-As maintains the diversity of large-ring motifs and fragments under pressure. Collectively, our results highlight how atomic-scale dihedral-angle flexibility, even among chemically similar elements, can influence medium-range structural diversity.

In terms of methodology, our study has demonstrated how automated MLIP fitting workflows enable the efficient generation of interatomic potentials for amorphous systems: in the present case of *a*-As, we have been able to generate a bespoke MLIP model that reproduces the experimental structure factor at a total computational cost of less than 50,000 CPU core hours. Beyond elemental systems, based on pilot studies for ternary chalcogenide memory materials in Ref. (*29*), we now expect that similar, detailed MLIP-driven studies of MRO can be carried out for multicomponent amorphous systems as well: for example, for Zr–Cu (*65*) or Co–V–Zr (*66*) metallic glasses. Applying automated MLIP generation broadly across different chemical families promises new insights into structure–property correlations in amorphous solids, offering a generalizable strategy to uncover how atomic-level interactions define medium-range order and macroscopic properties.

## METHODS

**MLIP generation with automated workflows**

The `autoplex` framework (version 0.0.7, available at https://github.com/autoatml/autoplex) was used to generate the RSS dataset in a largely automated manner (*29*). In each iteration, 10,000 random structures containing 6 to 24 atoms (even numbers) per cell were generated. Among them, 80% were constrained to possess between 2 and 8 symmetry operations, while the remaining 20% were generated without any symmetry constraints, to enhance the diversity of the initial structural pool.

Energies, forces, and stresses of the RSS-generated structures were obtained from DFT single-point computations with VASP (*67, 68*), managed by `atomate2` (*69*) through interfaces in `autoplex` (*29*). The DFT computations employed the projector augmented-wave (PAW) method (*68, 70*) and the r$^2$SCAN functional (*41*), a regularized form of SCAN (*71*) with improved numerical stability. The plane-wave energy cutoff was 600 eV. A *k*-point spacing of 0.2 Å$^{-1}$ was used to sample reciprocal space. Electronic self-consistency was converged to a tolerance of $1\times10^{-7}$ eV, and a Gaussian smearing width of 0.01 eV was used for partial occupancies. After each iteration, 100 data points were generated, with 90% used as the training set and the remaining 10% as the test set. The training and test sets for each iteration are cumulative, incorporating data from all previous iterations.

In addition to RSS, `autoplex` was also used to manage high-throughput DFT calculations for evaluating the energy of structures sampled from MD trajectories to refine the potentials. In this case, a workflow composed of the `DFTStaticLabelling` and `collect_dft_data` functions within `autoplex` was employed to generate a formatted dataset ready for MLIP training. For the larger cells of 216 atoms, we used Γ-point calculations, with all other parameters kept the same as those used for the RSS dataset.

**Hyperparameters**

In this work, iterative RSS was driven by GAP models (*22*, *32*, *33*). Except where noted, our GAP setup and hyperparameter choices match that of the earlier P-GAP-20 model (*45*). We disabled the "R6" baseline pair potential that had been used in Ref. (*45*), while retaining both the two-body term and the many-body SOAP term. The number of sparse points for the two-body term was 15 in both models, while that for SOAP was reduced from 8,000 (*45*) to 3,000 here, enabling faster evaluations at a speed of approximately 2 minutes per structure on a single CPU core. The weights for energies, forces, and stresses were assigned automatically within `autoplex`, based on the structures' distances from the energy convex hull [see also Ref. (*30*)].

We used the MACE framework (*35*) to retrain a model on the pure RSS dataset and fit refined datasets that were iteratively augmented with MD structures (Fig. 1A). For training MACE models, we used two message-passing layers, each with 128 channels. A correlation order of 3 was chosen, and spherical harmonics were set up to degree 3. The radial cutoff was 6 Å, resulting in a total receptive field of 12 Å per atom. The models were trained using the Huber loss function (*72*) for energies, forces, and stresses. Double precision was used throughout, and the maximum number of training epochs was set to 2,000. We trained all MACE models with the `mace-torch` package on an NVIDIA A100 80GB PCIe GPU, with the final complete training taking two and a half hours, resulting in an estimated energy consumption of ~0.75 kWh.

**MD simulations**

ML-driven MD simulations were performed with LAMMPS (*73*). The structure of *a*-As at ambient conditions was obtained via a melt–quench simulation protocol. An initial cubic cell containing 2,000 atoms [comparable to the 1,984 atoms used in models of *a*-P (*15*)], at an experimental density of 4.74 g cm$^{-3}$ (*37*), was created using `buildcell` (*31*, *74*) with hard-sphere potentials to ensure reasonable interatomic distances. The structures were first equilibrated at 3,000 K for 30 ps in the NVT ensemble to randomize atomic positions and remove structural ordering. This was followed by a melting simulation at 1,500 K for another 30 ps. The system was then cooled from 1,500 K to 300 K at a rate of $10^{12}$ K s$^{-1}$ (corresponding to 1.2 million MD steps). A comparison of different quench rates, and their effects on the resulting structures, can be found in the Supplementary Information. After reaching 300 K, the system was annealed for 50 ps in the NVT ensemble, followed by an additional 50 ps of annealing in the NPT ensemble using a Nosé–Hoover thermostat and barostat (*75*, *76*). The structure factor was calculated from the final 10 ps of the trajectory (200 snapshots) and compared with experimental data. Subsequently, compression simulations were performed at 300 K by gradually increasing the pressure from ambient conditions up to 6 GPa at a compression rate of 0.02 GPa ps$^{-1}$, consistent with that used in the previous study on *a*-P (*15*) for a fair comparison. The timestep in all MD simulations was 1 fs.

**Dihedral angles**

To calculate the dihedral angles in the dumbbell structural motifs (Fig. 2A), we first computed the sum of bond vectors from each central atom of connected trigonal units (*c* and *c'*) to its two neighboring atoms, yielding two directional vectors, **s** and **s'**. The dihedral angle, $\phi$, was then calculated as the angle between the planes spanned by {**s**, **r**$_{cc'}$} and {**s'**, **r**$_{cc'}$}, where **r**$_{cc'}$ is the vector from atom *c* to atom *c'*. To quantify the structural diversity encoded in the dihedral-angle distribution, we evaluated the information entropy $\mathcal{H}$ (also known as the Shannon entropy) using its standard definition from statistics and information theory (*60*) (Eq. 1).

**Void analysis**

We used a grid-based spatial search method to identify voids. In this method, a "void" grid was defined as a region of space where no atomic center existed within a specified cutoff distance. For each material, the cutoff distance was chosen to correspond to the first minimum of its radial distribution function: viz. 2.9 Å for *a*-As and 2.4 Å for *a*-P. A uniform Cartesian grid was generated within the simulation cell, with the grid spacing set to one-tenth of the chosen cutoff distance. Each grid point was then compared against the set of all atomic positions, including periodic images, with a cKDTree search (*77*); points with a nearest-atom distance greater than the cutoff were labelled as "void voxels". These void voxels were subsequently clustered via the DBSCAN algorithm using a clustering radius of 1.2 × grid spacing and a minimum cluster size of five points (*78*). Each cluster represented a contiguous free-volume region, with its volume equal to the number of voxels in the cluster multiplied by the voxel volume. The combined volume of all retained clusters was converted into an average equivalent spherical radius (Eq. 2).

Based on the recorded coordinates of each void voxel, PCA was applied to determine the direction along which the void spatial arrangement exhibits the greatest variance, namely the first principal component (PC1) axis (*79*). More specifically, given a set of $N_v$ voxel positions $\{\mathbf{x}_i\}_{i=1}^{N_v}$ in 3D space, the data were first mean-centered:

$$\mathbf{x}_i^c = \mathbf{x}_i - \bar{\mathbf{x}}, \tag{3}$$

where $\bar{\mathbf{x}}$ is the mean position of all void voxels. The covariance matrix was then computed as:

$$\mathbf{C} = \frac{1}{N_v - 1} \sum_{i=1}^{N_v} \mathbf{x}_i^c (\mathbf{x}_i^c)^T. \tag{4}$$

The first principal component (PC1) was obtained as the eigenvector $\mathbf{v}_1$ corresponding to the largest eigenvalue of $\mathbf{C}$. Each voxel position was then projected onto PC1 via

$$p_{i1} = (\mathbf{x}_i - \bar{\mathbf{x}}) \cdot \mathbf{v}_1 \tag{5}$$

where $p_{i1}$ denotes the coordinate of the *i*-th void along the PC1 axis. The coordinates along the second and third principal components (PC2 and PC3) can be similarly obtained by projecting $\mathbf{x}_i$ onto the corresponding eigenvectors.

**Acknowledgments**

This work was supported by UK Research and Innovation [grant number EP/X016188/1]. L.W. and J. G. were supported by ERC Grant MultiBonds (grant agreement no. 101161771; funded by the European Union. Views and opinions expressed are however those of the author(s) only and do not necessarily reflect those of the European Union or the European Research Council Executive Agency. Neither the European Union nor the granting authority can be held responsible for them.) S.R.E. is grateful to the Leverhulme Trust (UK) for a fellowship. We are grateful for computational support from the UK national high performance computing service, ARCHER2, for which access was obtained via the UKCP consortium and funded by EPSRC grant ref EP/X035891/1 [see also Ref. (*80*)]. J. G. and L.W. would like to acknowledge the Gauss Centre for Supercomputing e.V. (https:// www.gauss-centre.eu) by providing generous computing time on the GCS Supercomputer SuperMUC-NG at Leibniz Supercomputing Centre (www.lrz.de) (Project pn73da).

**Author contributions**

Y.L., S.R.E., and V.L.D. designed the study. Y.L. carried out the main computational work; R.A. carried out pilot studies with guidance from Y.Z. and additional input from Y.L.; L.W. carried out the bonding analyses. J.G. and V.L.D. supervised the computational work at Berlin and Oxford, respectively. Y.L., S.R.E., and V.L.D. wrote the manuscript, with input from L.W. and J.G., and all authors approved the final version.

**Data and materials availability**

Data supporting this work, including raw data and Python notebooks to reproduce the plots, will be made available via GitHub at `https://github.com/autoatml/papers-liu-as`. This work used the `autoplex` software (v0.0.7), which is openly available at `https://github.com/autoatml/autoplex`.

# Supplementary Information

# Medium-range structural order in amorphous arsenic

Yuanbin Liu[1], Yuxing Zhou[1], Richard Ademuwagun[1], Luc Walterbos[2],
Janine George[2,3], Stephen R. Elliott[4], Volker L. Deringer[1]*

[1]*Inorganic Chemistry Laboratory, Department of Chemistry, University of Oxford, Oxford OX1 3QR, UK*

[2]*Materials Chemistry Department, Federal Institute for Materials Research and Testing (BAM), Berlin, Germany*

[3]*Institute of Condensed Matter Theory and Solid-State Optics, Friedrich Schiller University Jena, Jena, Germany*

[4]*Physical and Theoretical Chemistry Laboratory, Department of Chemistry, University of Oxford, Oxford OX1 3QZ, UK*

\* volker.deringer@chem.ox.ac.uk

This PDF file includes supplementary text, Figs. S1–S10, and Table S1.

## Supplementary Text

For chemical-bonding analysis, we computed crystal orbital Hamilton population (COHP) (*S1*) and crystal orbital bond index (COBI) (*S2*) data with the program LOBSTER (*S3*, *S4*) and analyzed the output using LobsterPy (*S5*, *S6*).

LOBSTER analysis is based on a projection of the DFT wavefunctions onto an auxiliary basis of localized, atom-centered orbitals (4s and 4p on each As atom). Plotted alongside the electronic densities of states (DOS) in Fig. S6, the –COHP curves quantify bonding and antibonding interactions, while COBI provides information on the bond order across different energy levels. Gray As shows more metallic properties than black As and *a*-As, which has been attributed to its structural relationship with the simple cubic type (*S7*). Comparing *a*-As with the crystalline phases, the electronic structure of *a*-As appears to share the most similarities with that of the black-As phase. This is again in line with previous observations (*S8*). However, whereas black As has a small bandgap (0.3 eV), our *a*-As model appears to feature states at the Fermi level. We think that this is likely caused by defect states due to under-coordinated As atoms, consistent with an earlier study of the electronic properties of *a*-P (*S9*).

The electronic-structure analyses was performed at the r$^2$SCAN+rVV10 level of theory (*S10*), with a plane-wave energy cutoff of 600 eV. For crystalline structures, a *k*-point spacing of 0.2 Å$^{-1}$ was employed, while for *a*-As we used Γ-point calculations. For the experimentally known crystalline As phases (gray and black), the initial structures were taken from the Materials Project (*S11*). For the hypothetical modifications (yellow, violet, and fibrous), the initial structures were generated by substituting As into the corresponding P structures (*S9*). (We note that other crystalline structures for As have been discussed based on experiments (*S12*) and computation (*S8*, *S13*), but we do not include those in Fig. S6.) All relevant, initial crystalline structures were subsequently fully relaxed using the r$^2$SCAN+rVV10 functional. The amorphous structure contained 500 atoms and was generated by relaxing the melt–quench configuration with our MLIP. The nearest-neighbor bonds were identified using the `COHPgenerator` keyword in LOBSTER, with the cut-off for bond detection set to 2.9 Å.

# Supplementary Figures

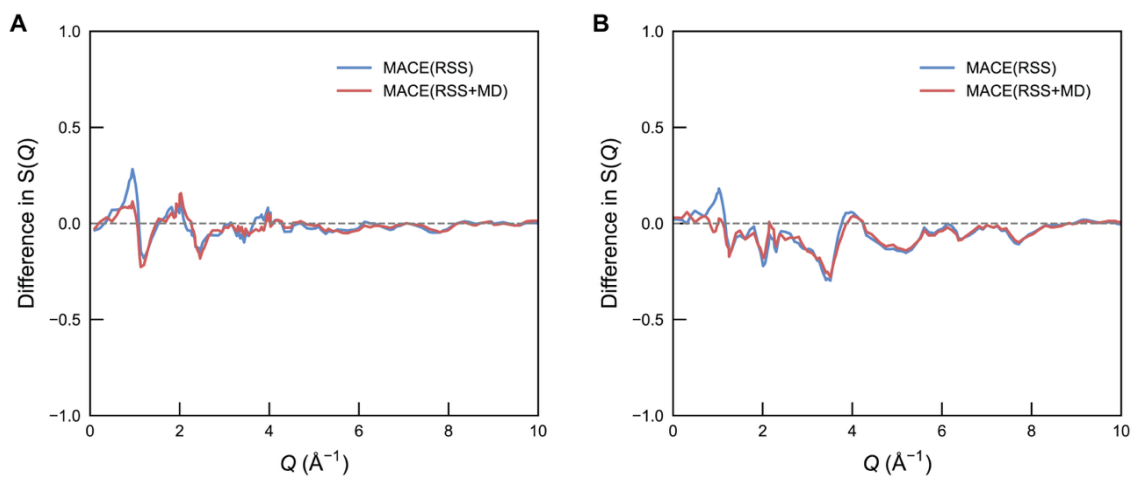

**Fig. S1 Structure-factor differences between experiments and two MACE models.** (**A**) Experimental data from Smith et al. (*S14*). (**B**) Experimental data from Bellissent and Tourand (*S15*). The curves show the difference obtained by subtracting the simulated values from the experimental data for each model. It shows that the additional MD-based refinement primarily increases the intensity of the predicted FSDP, whereas the remainder of the curve is essentially unchanged compared to that predicted by the purely RSS-trained model.

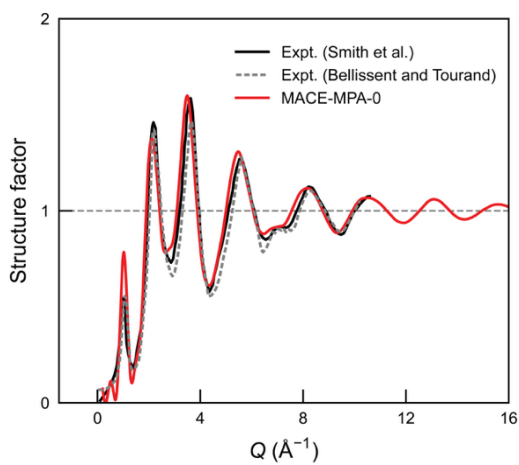

**Fig. S2 Benchmarking the MACE-MPA-0 foundation model against experimental data.** The structure factor of *a*-As calculated using the MACE-MPA-0 model (*S16*) shows qualitative agreement with experimental results from Smith et al. (*S14*) as well as Bellissent and Tourand (*S15*). However, the model exhibits quantitative discrepancies, notably overestimating the intensity of the first sharp diffraction peak (FSDP).

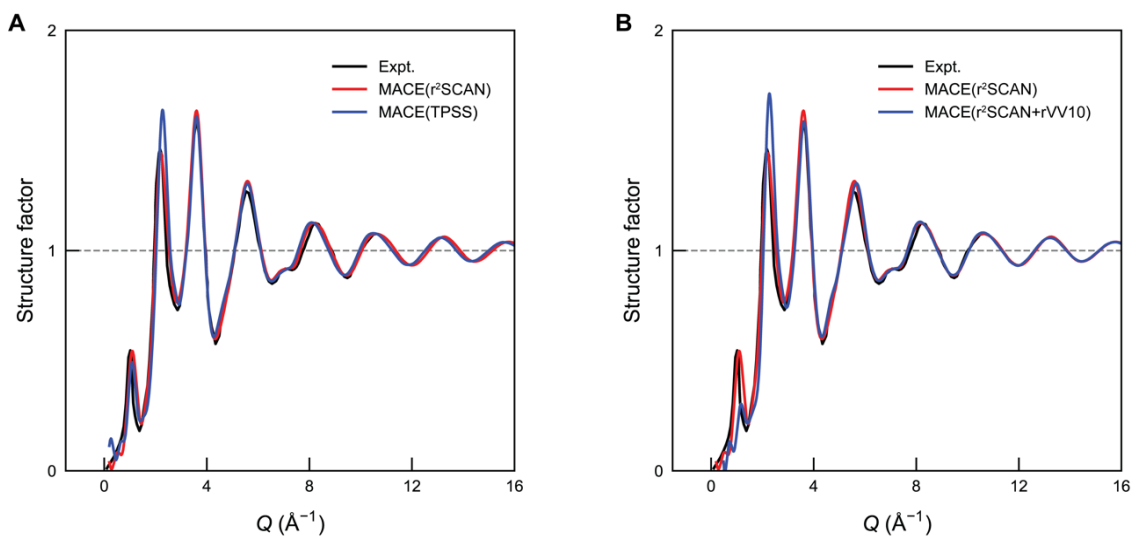

**Fig. S3 Benchmarking meta-GGA functionals via structure factors of amorphous structures.** Structure factors of models of *a*-As obtained from potentials trained with TPSS (*S17*), r$^2$SCAN (*S18*), and r$^2$SCAN+rVV10 (*S10*) functionals are compared with experimental data from Ref. (*S14*). (**A**) Comparison of r$^2$SCAN and TPSS; (**B**) comparison of r$^2$SCAN and r$^2$SCAN+rVV10. The best agreement is obtained with r$^2$SCAN, while TPSS yields intermediate results, overestimating the intensity of the second peak. In contrast, r$^2$SCAN+rVV10 significantly underestimates the intensity of the FSDP and overestimates the intensity of the second peak.

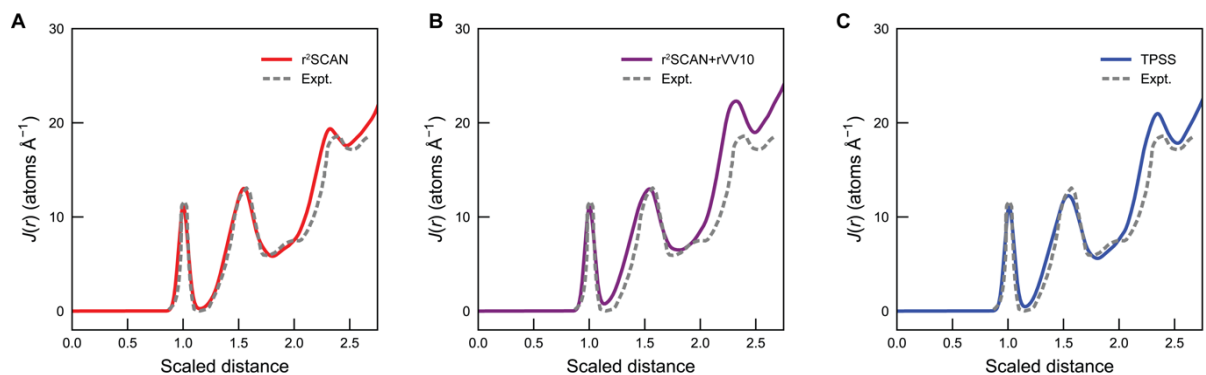

**Fig. S4 Benchmarking meta-GGA functionals via radial distribution functions of amorphous arsenic.** Radial distribution functions $J(r)$ calculated with different meta-GGA functionals are compared with experimental data (*S19*). We show results based on (**A**) r$^2$SCAN, (**B**) r$^2$SCAN+rVV10, and (**C**) TPSS. All functionals reproduce the experimental trends, with r$^2$SCAN showing the closest agreement and further validating the accuracy of our choice for functional.

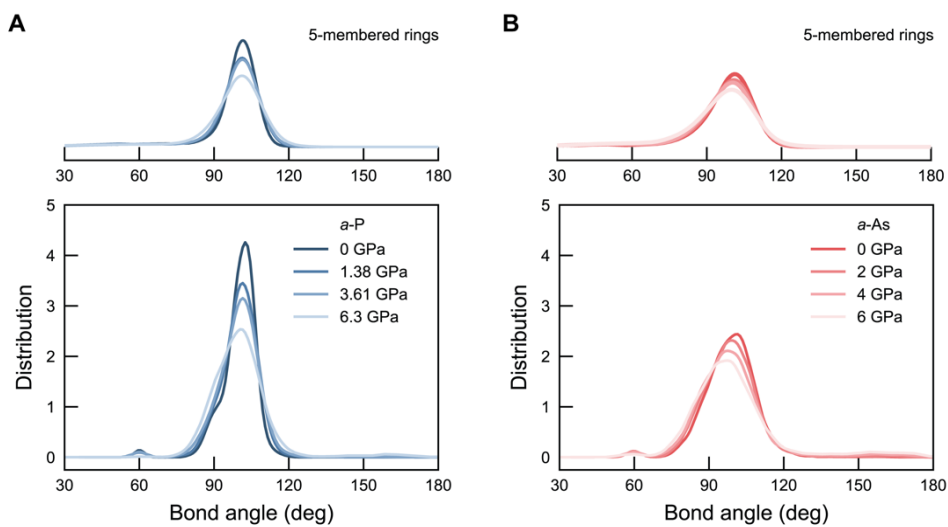

**Fig. S5 Bond-angle distribution as a function of pressure.** (**A**) *a*-P and (**B**) *a*-As. Kernel density estimates at the top of each panel show the bond-angle distributions within 5-membered rings, while the lower panels display the probability-density functions of the total bond-angle distributions calculated for all atoms in the systems. Notably, the dominant peak in the total bond-angle distribution coincides with that of the 5-membered rings, suggesting that their similar angular preferences facilitate the formation of 5-membered rings in both *a*-P and *a*-As.

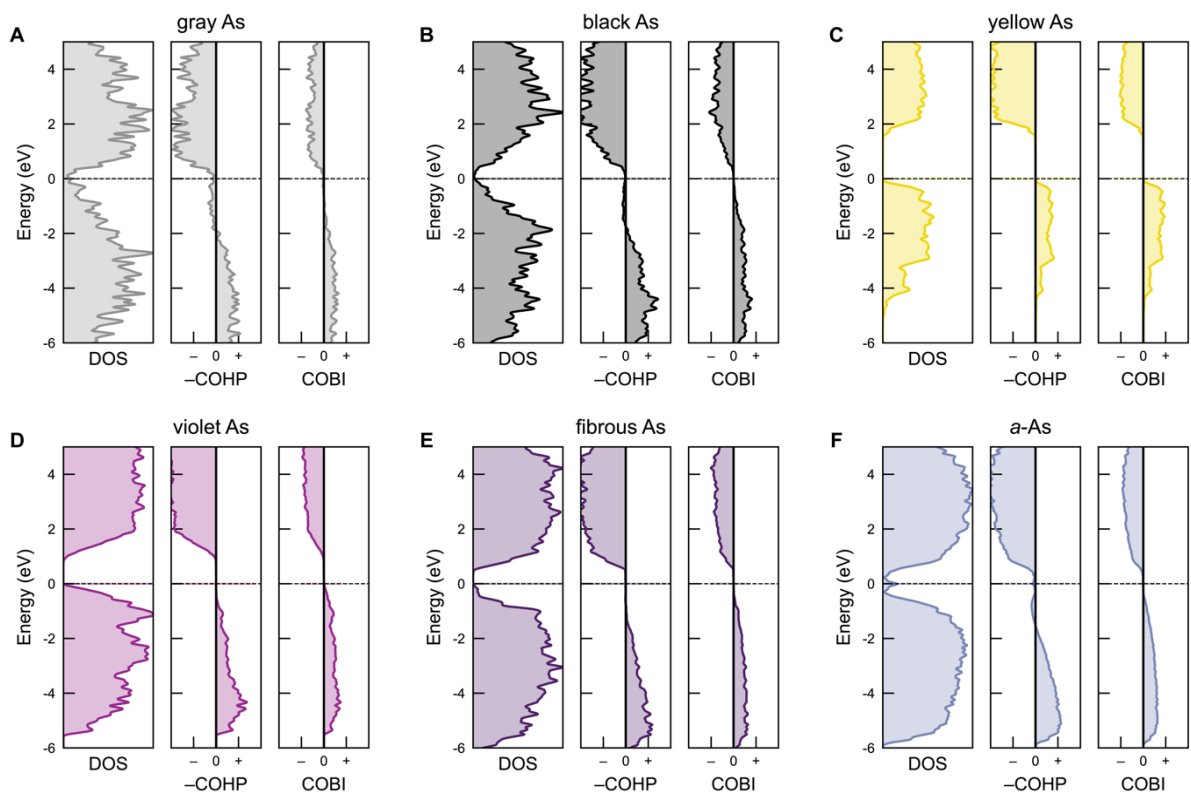

**Fig. S6 Electronic structure and chemical-bonding analysis of crystalline and amorphous As phases.** Electronic density of states (DOS), projected crystal orbital Hamilton population (COHP), and crystal orbital bond index (COBI) plots of six relevant As phases: (**A**) gray As, (**B**) black As, (**C**) yellow As, (**D**) hypothetical "violet As" (derived from violet P by elemental substitution and subsequent relaxation), (**E**) as in (D) but for "fibrous As", and (**F**) amorphous As (*a*-As). See the Supplementary Text on p. S2 for details.

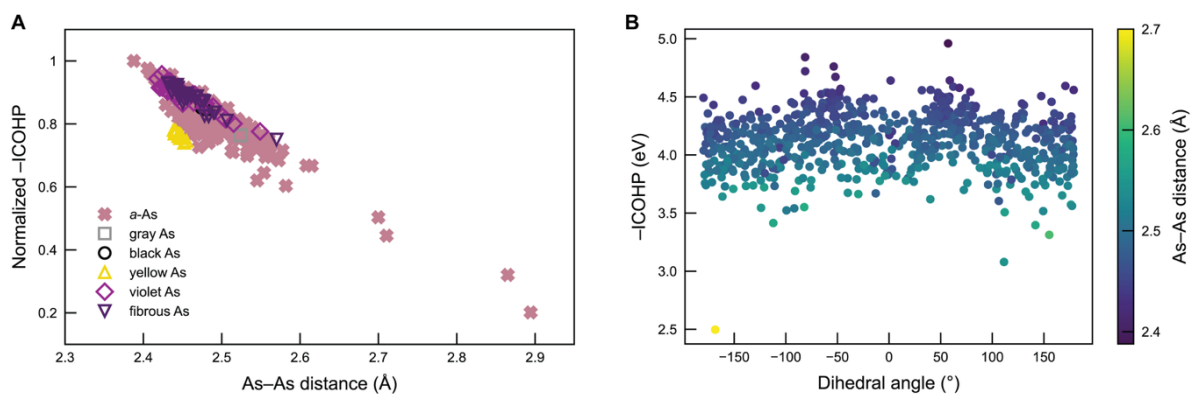

**Fig. S7 Structure and bonding in crystalline and amorphous As.** (**A**) Bond length–bond strength relationship, where each symbol represents an As–As bond. The ICOHP values, taken to measure the bond strength, are normalized to the value for the strongest bond in the *a*-As network (–ICOHP = 4.95987 eV). Our analysis shows that the bond strengths of the nearest neighbour interactions in *a*-As span a comparably large range. The average bond energy (–ICOHP = 4.10231 eV) and length in *a*-As are roughly equal to the average of the bond energy and length in black (–ICOHP = 4.20101 eV). (**B**) Dependence of bond strength on dihedral angle in *a*-As. The color bar denotes the As–As bond distance, illustrating that stronger bonds (more negative ICOHP values) are generally associated with shorter bond lengths, whereas the dihedral angle shows no strong correlation with the bond strength, as indicated by a Spearman coefficient (*S20*) of 0.0057. The electronic-structure computations were performed using settings as described in the Supplementary Text (p. S2).

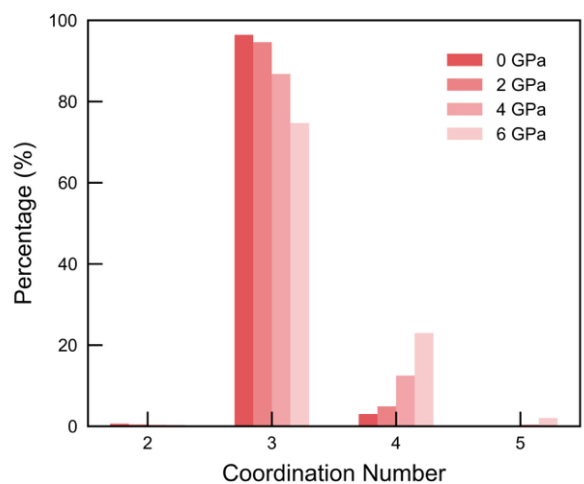

**Fig. S8 Coordination-number distribution of *a*-As under pressure.** The proportion of 3-fold coordinated atoms decreases with increasing pressure, while the occurrence of over-coordinated environments (primarily 4-fold) becomes more prominent. The 2-fold coordination remains negligible for all pressures. A cutoff distance of 2.9 Å was employed to define coordination numbers.

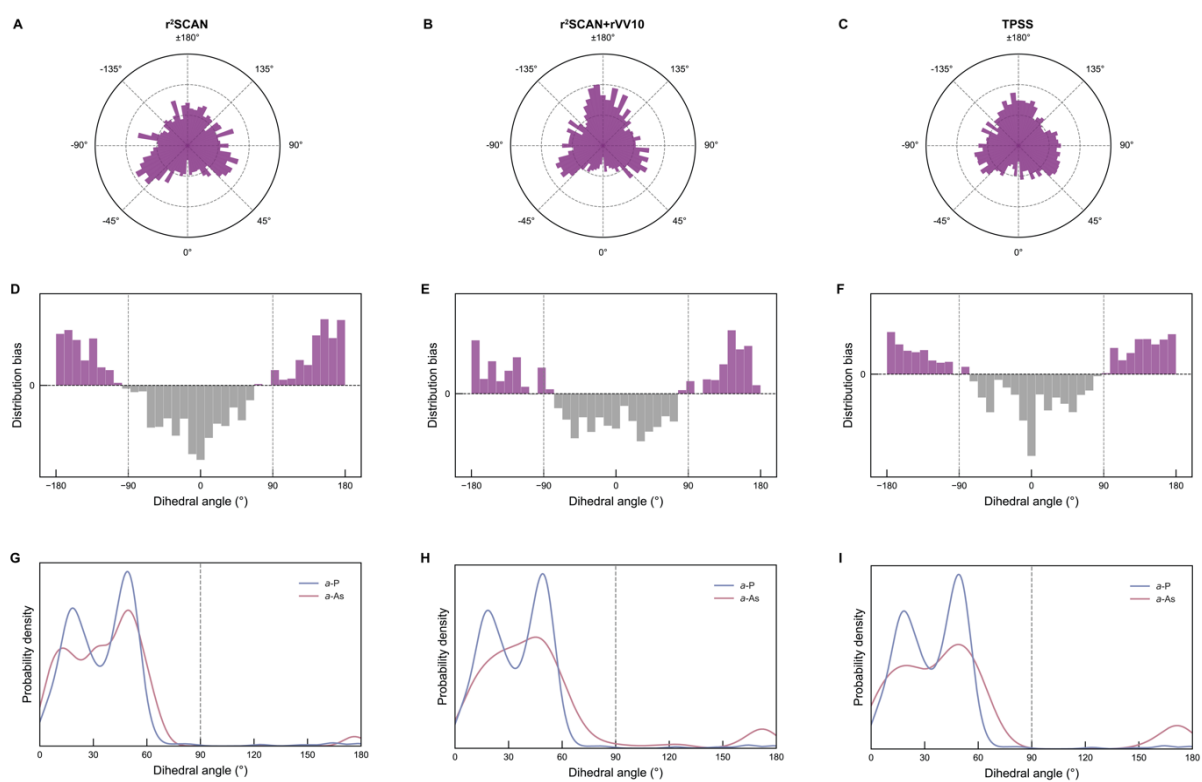

**Fig. S9 Dihedral-angle characteristics of models of *a*-As simulated using MLIPs trained with different functionals.** (**A–C**) Polar histograms of dihedral angle distributions for models of *a*-As obtained from MLIPs trained with r$^2$SCAN, r$^2$SCAN+rVV10, and TPSS, respectively. (**D–F**) Distribution bias in dihedral angles between large ($n > 6$) and small ($n \leq 6$) membered rings for the corresponding functionals. (**G–I**) Probability-density distributions of dihedral angles in $X_3]X_2[X_3$ fragment clusters, the most prevalent cluster type in *a*-As, for the corresponding functionals. It indicates that changing the functional does not affect the main conclusions regarding dihedral angles: the dihedral-angle distribution in *a*-As is broad; larger rings in *a*-As tend to comprise large dihedral angles (>90°); and dihedral angles in $X_3]X_2[X_3$ clusters are much more likely to be smaller than 90°.

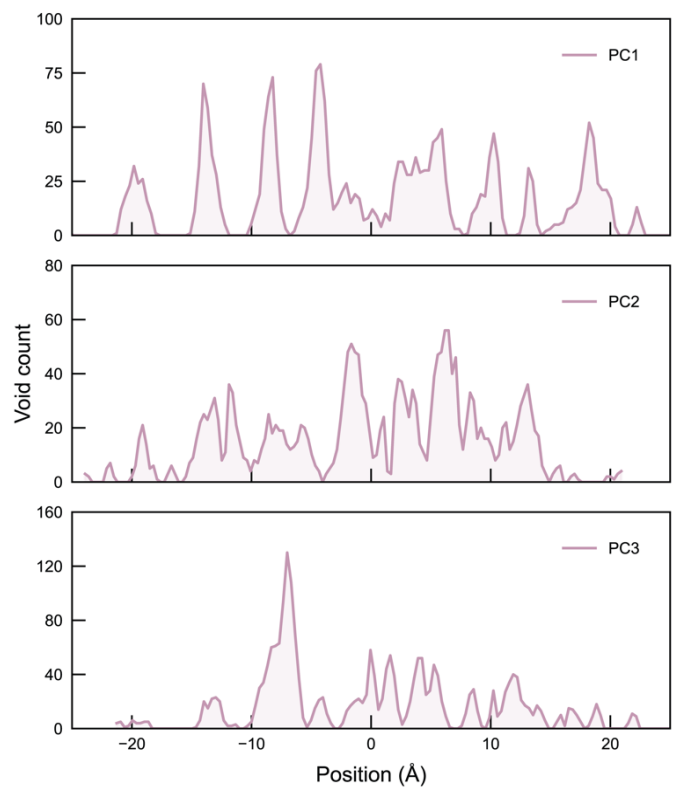

**Fig. S10 Spatial distributions of voids projected along principal-component axes.** The three panels correspond to the distributions along the first (PC1), second (PC2), and third (PC3) principal components. This plot illustrates that peak patterns are more distinct along PC1.

S12

# Supplementary Table

**Table S1 Comparison of lattice constants and relative errors for the gray and black allotropes of As, computed using various DFT functionals.** All relative errors are computed with respect to experimental values. In the case of r$^2$SCAN, we note a strong overestimation of the c lattice parameter of gray As (marked by * in the table), which appears to reflect a missing description of van der Waals attraction between the layers. For gray As, we tested the effect of different $k$-point mesh densities, ranging from 4×4×2 to 10×10×3 and 20×20×6, corresponding to $k$-point spacing of 0.5 Å$^{-1}$, 0.2 Å$^{-1}$, and 0.1 Å$^{-1}$, respectively. Interestingly, the coarsest grid provides the closest prediction of the $c$-axis with a relative error of 1.3%, whereas the finer grids (0.2 Å$^{-1}$ and 0.1 Å$^{-1}$) yield similar results with much larger deviations (~45%), suggesting that numerical noise in r$^2$SCAN may be smoothed out by a coarser mesh.

|  | Methods | Lattice parameters (Å) | | | Relative errors (%) | | |
| --- | --- | --- | --- | --- | --- | --- | --- |
|  |  | a | b | c | a | b | c |
| Gray As | Experiment (*S21*) | 3.760 | 3.760 | 10.547 | — | — | — |
|  | r$^2$SCAN | 3.588 | 3.588 | 15.338* | –4.6 | –4.6 | 45.4* |
|  | r$^2$SCAN+rVV10 | 3.775 | 3.775 | 10.276 | 0.4 | 0.4 | –2.6 |
|  | TPSS | 3.803 | 3.803 | 10.635 | 1.1 | 1.1 | 0.8 |
|  | PBE | 3.826 | 3.826 | 10.645 | 1.8 | 1.8 | 0.9 |
|  | PBE-D3 | 3.794 | 3.794 | 10.058 | 0.9 | 0.9 | –4.6 |
| Black As | Experiment (*S22*) | 3.620 | 10.850 | 4.480 | — | — | — |
|  | r$^2$SCAN | 3.650 | 11.387 | 4.740 | 0.8 | 4.9 | 5.8 |
|  | r$^2$SCAN+rVV10 | 3.671 | 10.941 | 4.499 | 1.4 | 0.8 | 0.4 |
|  | TPSS | 3.727 | 11.050 | 4.473 | 3.0 | 1.8 | –0.2 |
|  | PBE | 3.712 | 11.491 | 4.657 | 2.5 | 5.9 | 4.0 |
|  | PBE-D3 | 3.746 | 10.841 | 4.289 | 3.5 | –0.1 | –4.2 |

# Supplementary References